\documentclass[12pt]{iopart}
\usepackage{psfig}
\usepackage{epsfig}

\newcommand{\be}{\begin{equation}}
\newcommand{\ee}{\end{equation}}
\def\(#1){(\ref{#1})}
\newcommand{\bea}{\begin{eqnarray}}
\newcommand{\eea}{\end{eqnarray}}

\newcommand{\deriv}[1]{{\partial\over\partial{#1}}}
\newcommand{\lav}{\left\langle}
\newcommand{\rav}{\right\rangle}
\newcommand{\eg}{{\it e.g.}}
\newcommand{\ie}{{\it i.e.}}
\newcommand{\av}[1]{\left\langle #1 \right\rangle}

\newcommand{\tw}{t_{\rm w}}
\newcommand{\dt}{t\!-\!t_{\rm w}}
\newcommand{\twlim}{t_{\rm w}\to\infty}
\newcommand{\tlim}{t\to\infty}
\newcommand{\mw}{m_{\rm w}}
\newcommand{\Ew}{E_{\rm w}}
\newcommand{\distrm}{\sigma(m|E)}
\newcommand{\trapmean}{\overline{m}(E)}

\newcommand{\trapvar}{\Delta^2(E)}
\newcommand{\globalmbr}[1]{\left\langle m(#1) \right\rangle}

\newcommand{\Emprobc}{P(E,m,t|E_{\rm w},m_{\rm w},t_{\rm w})}
\newcommand{\Ct}{\tilde{C}}
\newcommand{\Rt}{\tilde{\chi}}
\newcommand{\C}{C(t-t_{\rm w})}
\newcommand{\R}{\chi(t-t_{\rm w})}
\newcommand{\CC}{C(t,t_{\rm w})}
\newcommand{\RR}{\chi(t,t_{\rm w})}
\newcommand{\XX}{X(t,t_{\rm w})}
\newcommand{\teff}{T_{\rm eff}}
\newcommand{\tg}{T_{\rm g}}
\newcommand{\eq}[1]{~(\ref{#1})}
\newcommand{\xih}{\xi_h}
\newcommand{\tauh}{\tau_h}

\begin{document}

\title[Effective temperatures in non-mean field systems]{Fluctuation-dissipation relations and effective temperatures in
simple non-mean field systems}

\author{Peter Sollich\dag\footnote[3]{%
Email: peter.sollich@kcl.ac.uk}
, Suzanne Fielding\ddag\footnote[4]{%
Present address: Department of Physics and Astronomy $\&$ Polymer IRC,
University of Leeds, Leeds, LS2 9JT, U.K.}
and Peter Mayer\dag\footnote[5]{%
Email: peter.mayer@kcl.ac.uk}
}

\address{\dag\ Department of Mathematics, King's College London,
Strand, London, WC2R 2LS, U.K.}

\address{\ddag\ Department of Physics and Astronomy, University of Edinburgh,
Mayfield Road, Edinburgh, EH9 3JZ, U.K.}

\begin{abstract}
We give a brief review of violations of the fluctuation-dissipation
theorem (FDT) in out-of-equilibrium systems; in mean field scenarios
the corresponding fluctuation-dissipation (FD) plots can, in the limit
of long times, be used to define a {\em effective temperature} $\teff$
that shares many properties of the true thermodynamic temperature $T$. We
discuss carefully how correlation and response functions need to be
represented to obtain meaningful limiting FD plots in {\em non-mean
field} systems. A minimum requirement on the resulting effective
temperatures is that they should be independent of the observable
whose correlator and response are being considered; we show for two
simple models with glassy dynamics (Bouchaud's trap model and the
Glauber-Ising chain at zero temperature) that this is generically not
the case.  Consequences for the wider applicability of effective
temperatures derived from FD relations are discussed; one intriguing
possibility is that at least the limit of the FDT violation factor for
well separated times may generically be observable-independent and so
could yield a meaningful $\teff$.
\end{abstract}



\section{Introduction: Fluctuation-dissipation relations and effective
temperatures} 

One of the core ideas of statistical mechanics is that {\em
equilibrium} states can be accurately described in terms of only a
small number of thermodynamic variables, such as temperature and
pressure.  For glassy systems, which can remain far from equilibrium
on very long time scales, no similar simplification exists {\em a
priori}; the whole past history of a sample is in principle required
to specify its state at a given time. This complexity makes the
theoretical analysis of glasses very awkward, and one is driven
instead to look for a description of out-of-equilibrium states in
terms of a few effective thermodynamic parameters. The focus of the
present paper is on one such parameter, the {\em effective
temperature}; this can be defined on the basis of fluctuation-dissipation
(FD) relations between correlation and response functions and has
proved to be very fruitful in mean field systems~\cite{CugKurPel97}.

The use of FD relations to quantify the out-of-equilibrium dynamics in
glassy systems is motivated by the occurrence of {\em
ageing}~\cite{BouCugKurMez98}: The time scale of response to an
external perturbation increases with the age (time since preparation)
$\tw$ of the system. As a consequence, time translational invariance
(TTI) and the equilibrium fluctuation-dissipation
theorem~\cite{Reichl80} (FDT) relating correlation and response
functions break down. To quantify this, consider the autocorrelation
function for a generic observable $m$ of a system, defined as
\be
\CC=\lav m(t)m(\tw)\rav -\lav m(t) \rav \lav m(\tw) \rav
\label{corr}
\ee
The associated `impulse response' function can be defined as
\[
R(t,\tw)=\left.\frac{\delta \lav m(t)\rav}{\delta h(\tw)}\right|_{h=0}
\]
and gives the linear response of $m(t)$ to a small impulse in its conjugate
field $h$ at time $\tw$. An equivalent way of characterizing the
linear response is via the `step response' function
\be
\RR=\int_{\tw}^t\!
dt'\, R(t,t')
\label{switch_on}
\ee
which tells us how $m$ responds to a small step $h(t)=h\Theta(\dt)$ in the
field. In spin glasses, this response function would be called the
zero-field cooled response because the field is only switched on a
certain `waiting time' $\tw$ after preparation of the system (\eg\
by cooling) at time $t=0$.

Now, in {\em equilibrium}, $\CC=\C$ by TTI (similarly for $R$ and
$\chi$), and the FDT reads
\be
-\deriv{\tw} \R = R(t,\tw) = \frac{1}{T}\deriv{\tw}\C
\label{eq_fdt}
\ee
with $T$ the thermodynamic temperature (we set $k_{\rm B}=1$). A
parametric `FD plot' of $\chi$ vs.\ $C$ is
thus a straight line of slope $-1/T$. In the ageing case, the
violation of FDT can be measured by an FD ratio, $\XX$, defined
through~\cite{CugKur93,CugKur94}
\be
\label{eqn:non_eq_fdt}
-\deriv{\tw} \RR= R(t,\tw) = \frac{{\XX}}{T}\deriv{\tw}\CC
\ee
In equilibrium, due to TTI, the derivatives $\partial/\partial \tw$ in
the FDT~(\ref{eq_fdt}) could equally be replaced by
$-\partial/\partial {t}$; one could therefore argue that a similar
replacement could be made in\eq{eqn:non_eq_fdt}, leading to a
different definition of $\XX$. However, the latter would make rather
less sense, since in a situation without TTI only the $\tw$-derivative
of $\RR$ is directly related to the impulse response
$R(t,\tw)$. Adopting therefore the definition\eq{eqn:non_eq_fdt}, one
sees that values of $X$ different from unity mark a violation of
FDT. In glasses, these can persist even in the limit of long times,
indicating strongly non-equilibrium behavior even though one-time
observables of the system---such as entropy and average energy---may
have already settled to essentially stationary values.

Remarkably, the FD ratio for several {\em mean field}
models~\cite{CugKur93,CugKur94} assumes a special form at long times:
Taking $\tw\to\infty$ and $t\to\infty$ at constant $C=C(t,\tw)$,
$X(t,\tw)\to X(C)$ becomes a (nontrivial) function of the single
argument\footnote{%
This definition is not quite as stringent as requiring that $X(t,\tw)$
be a function of $C(t,\tw)$ only, at long times $\tw$ and $t$. For
some coarsening
models at criticality one finds, for example, that $X(t,\tw)=f(\tw^a
C(t,\tw))$ for long times, in terms of a scaling function $f(\cdot)$ and
a positive exponent $a$~\cite{GodLuc_ferro}; $X$ is therefore not a
function of $C$ alone. Nevertheless, making $t$ and $\tw$ large at
constant $C$, one finds a nontrivial function $X(C)$ which in this
case is discontinuous, $X(C)=f(\infty)$ for $C>0$ and $X(C)=f(0)$ for
$C=0$.
}
$C$. If the equal-time correlator $C(t,t)$ also approaches a constant
$C_0$ for $t\to\infty$, it follows that
\be
\chi(t,\tw)=\frac{1}{T}\int_{C(t,\tw)}^{C_0}
\!dC\, X(C)
\label{mf_limit}
\ee
Graphically, this limiting non-equilibrium FD relation is obtained by
plotting $\chi$ vs $C$ for increasingly large times; from the slope
$-X(C)/T$ of the limit plot, an {\em effective
temperature}~\cite{CugKurPel97} can be defined as $\teff(C)=T/X(C)$.

In the most general ageing scenario, a system displays dynamics on
several characteristic time scales, one of which may remain finite as
$\twlim$, while the others diverge with $\tw$.  If, due to their
different functional dependence on $\tw$, these time scales become
infinitely separated
as $\twlim$, they form a set of
distinct `time sectors'; in mean field, $\teff(C)$ can then be shown
to be {\em constant} within each such sector~\cite{CugKur94}. In the short
time sector ($\dt=O(1)$), where $C(t,\tw)$ decays from $C_0$ to some
plateau value,
one
generically has quasi-equilibrium with $\teff=T$, giving an initial
straight line with slope $-1/T$ in the FD plot. The further decay of
$C$ (on ageing time scales $\dt$ that grow with $\tw$) gives rise to
one of three characteristic shapes: (i) In models which statically
show one step replica symmetry breaking (RSB), \eg\ the spherical
$p$-spin model~\cite{CugKur93}, there is only one ageing time sector
and the FD plot exhibits a second straight line, with $\teff>T$. (ii)
In models of coarsening and domain growth, \eg\ the $O(n)$ model at
large $n$, this second straight line is flat, and hence
$\teff=\infty$~\cite{CugDea95}. (iii) In models with an infinite
hierarchy of time sectors (and infinite step RSB in the statics, \eg\
the SK model) the FD plot is instead a continuous
curve~\cite{CugKur94}. 

$\teff$ has been interpreted as a time scale dependent non-equilibrium
temperature, and within mean field has been shown to display many of
the properties associated with a thermodynamic
temperature~\cite{CugKurPel97}. For example (within a given
time sector), it is the reading which would be shown by a thermometer
tuned to respond on that time scale. Furthermore---and of crucial
importance to its interpretation as a temperature---it is independent
of the observable\footnote{%
This applies to a wide class of observables, but has not to our
knowledge been established in complete generality. The restriction
arises because the observables that are commonly used in mean field
models are defined
in terms of random fields or couplings, and are therefore uncorrelated
with the energy of the system. It is possible that for other variables,
which {\em are} correlated with the energy, different effective
temperatures would be obtained (S Franz and F Ritort, private
communication). This effect is observed explicitly in oscillator
models with Monte Carlo dynamics~\protect\cite{Nieuwenhuizen00b}.
} $m$ used to construct the FD
plot~\cite{CugKurPel97}. 

While the above picture is well established in mean field, its status
in non-mean field models is less obvious. To check its validity, one
must demonstrate a) that a limiting FD plot exists, b) that it gives
effective temperatures that are independent of the observable-field
pair used to calculate $C$ and $\chi$, and c) that the effective
temperature is constant within a given time sector, \ie\ the same for
all time scales differing only by factors of order unity. One then
expects FD plots similar to those for mean field systems, composed
either of a set of straight lines, or of a continuous curve in the
case of an infinite number of distinct ageing time sectors. Of course,
full independence of the effective temperature from the observable
considered may be too strong a requirement for non-mean field systems,
but one would at least expect independence within a large class of
sufficiently `neutral' observables; we return to this point in
Sec.~\ref{sec:discussion}.

Encouragingly, molecular dynamics (MD) and Monte Carlo (MC)
simulations of binary Lennard-Jones mixtures~\cite{KobBar00}, as well
as MC simulations of frustrated lattice gases~\cite{AreRicSta00}
(which loosely model structural glasses, whose phenomenology is
similar to that of the $p$-spin model) show a limiting plot of type
(i). Oscillator models with MC dynamics show similar plots, but the
initial short-time regime where one has effective equilibrium is
absent and one finds only a single straight line~\cite{BonPadRit98}.
MC simulations of the Ising model with conserved and
non-conserved order parameter in dimension $d=2$ and 3 show a plot of
type (ii)~\cite{Barrat98}. And MC simulations of the Edwards-Anderson
model in $d=3$ and 4~\cite{MarParRicRui98} give a plot of type
(iii). The majority of existing studies, however, do not show that
$\teff$ is independent of observable (notable exceptions being
Refs.~\cite{KobBar00,AreRicSta00}) since they consider just one
observable-field pair. In the Backgammon model (see \eg\
Ref.~\cite{GodLuc_urn} for a summary of results), several observables
have recently been considered and differences in the FD ratios found;
however, these differences only occur in sub-leading terms which decay
to zero at long times, and so it is not clear how seriously
this observable-dependence needs to be taken.

Below, we therefore give some results for two models for which FD
relations can be calculated for a broad range of observables; our aim
is to clarify the status of FD-derived effective temperatures in
non-mean field systems. First, however, we pause briefly to discuss
the appropriate representation of FD plots in such systems. As
explained above, for mean field systems the existence of a limiting
relation\eq{mf_limit} between response $\chi$ and correlation $C$
ensures that parametric plots of $\chi$ versus $C$ converge, for long
times, to a limiting FD plot whose negative slope directly gives
$X(C)/T$. Eq.\eq{mf_limit} implies that the plots can be produced
either with $t$ as the curve parameter, holding the earlier time $\tw$
fixed, or vice versa. Since it is normally easier to fix the time
$\tw$ at which a field is switched on and measure the resulting
response as a function of $t$---rather than observing the response at
fixed $t$ to fields switched on at a range of earlier times
$\tw$---the first version is the one that is conventionally
used~\cite{CugKur93,CugKur94}.

In general, however, it is clear that the
definition~(\ref{eqn:non_eq_fdt}) ensures a slope of $-\XX/T$ for a
parametric $\chi$-$C$ plot {\em only} if $\tw$ is used as the
parameter, with $t$ being fixed. This issue becomes particularly
important when---for the particular observable chosen---the equal-time
correlator $C(t,t)$ diverges or decreases to zero for $t\to\infty$,
instead of approaching a constant value. A series of `raw' FD plots at
increasing $t$ will then either grow or shrink in scale indefinitely,
preventing the existence of a limiting FD plot for $t\to\infty$. It
therefore appears sensible to {\em normalize} the FD plot. The
normalization factor must be the same for the $\chi$-axis and the
$C$-axis and must be independent of $\tw$, in order to preserve the
connection between $X$ and the slope of the plot. To `attach'
the plot to a definite point on the $C$-axis, it is thus sensible to
normalize by $C(t,t)$, plotting $\Rt(t,\tw)=\chi(t,\tw)/C(t,t)$ vs
$\Ct(t,\tw)=C(t,\tw)/C(t,t)$. In this representation, a limiting FD
plot {\em may} be approached as $t$ is increased. If this is the case, then
$X(t,\tw)$ converges to a nontrivial limit for large times, but now at
a constant value of the {\em normalized} correlator $\Ct$ (rather than
of $C$ itself). Also, {\em if a limiting plot exists}, it can be obtained
from plots of $\Rt$ vs $\Ct$ that use {\em either} $t$ or $\tw$ as
the curve parameter; the normalization still has to be performed with
$C(t,t)$ and not with $C(\tw,\tw)$, however, in order to retain the
link between $X$ and the slope of the limit plot.

One final benefit of the method of producing FD plots discussed
above---normalize by $C(t,t)$, and use $\tw$ as the curve
parameter---is that the plots for the two types of response function
most frequently considered are trivially related. We have worked above
with the `switch-on' (or zero-field cooled) response
function\eq{switch_on}; a common alternative is the `switch-off' (or
thermo-remanent) response, defined for a field switched on at time $0$
and off again at time $\tw$:
\be
\chi_-(t,\tw)=\int_0^{\tw}\!
dt'\, R(t,t') = \chi(t,0)-\chi(t,\tw)
\label{switch_off}
\ee
With $\tw$ being used as the curve parameter, the FD plots of $\chi$
vs $C$ and $\chi_-$ vs $C$ are then trivially related (by a reflection
about the horizontal axis and a vertical shift); the same remains true
after normalization of the curves by $C(t,t)$. If, on the other hand,
one were to follow the naive approach of using $t$ as the curve
parameter---keeping $\tw$ fixed---and normalizing by $C(\tw,\tw)$,
then the plots for $\chi$ and for $\chi_-$ are not related in any
obvious way and can in fact have very different
shapes~\cite{fielding_thesis}.

We end this section with a brief comment on a recent
proposal~\cite{Rocco_SW} to construct FDT plots from the response and
the {\em disconnected} correlation function
\[
C_{\rm dis}(t,\tw) = \lav m(t)m(\tw) \rav
\]
as opposed to the {\em connected} one\eq{corr} used above. This can
avoid the normalization problem in some cases: If, for example, one is
dealing with a systems of spins $s_i=\pm 1$, and $m$ is a staggered
magnetization with a random field, then $C_{\rm dis}$ reduces to a
spin-spin autocorrelation function and so is automatically normalized
at equal times, $C_{\rm dis}(t,t)=1$. While apparently convenient, use
of $C_{\rm dis}$ in FD plots appears to us to have little theoretical
justification. A serious drawback is that FD plots constructed from
$C_{\rm dis}$ are no longer invariant under simple affine
transformations of the observable, $m' = am + b$. This seems to us to
violate the physical intuition that such trivially related observables
should not be able to reveal significantly different glassy
physics. We would therefore advocate use of the connected correlation
function\eq{corr} throughout; this ensures that affine transformations
of observables, as defined above, leave all FD plots invariant (up to
a trivial overall scale factor of $a^2$). Of course, if the observable
is such that it has no drift, \ie\ $\lav m(t)\rav =$ constant, then
disconnected or connected correlation functions can be used
interchangeably, since they differ only by a trivial additive
constant. This may be the case for sufficiently `neutral' observables
(or could in fact be part of the criterion for deciding whether an
observable is neutral); see the discussion in
Sec.~\ref{sec:discussion}.

\section{Model system 1: Bouchaud's trap model}

Trap models~\cite{Bouchaud92,MonBou96,RinMaaBou00} are obvious
candidates for a study of FD violation in systems with glassy
dynamics. Popular as alternatives to the microscopic spin models
discussed above, they capture ageing within a simplified single
particle description. The simplest such model~\cite{Bouchaud92}
comprises an ensemble of uncoupled particles exploring a spatially
unstructured landscape of (free) energy traps by thermal
activation. The tops of the traps are at a common energy level and
their depths $E$ have a `prior' distribution $\rho(E)$ ($E>0$).  A
particle in a trap of depth $E$ escapes on
a time scale $\tau(E)=\tau_0\exp({E}/{T})$ and hops into
another trap, the depth of which is drawn at random from
$\rho(E)$.  The probability, $P(E,t)$, of finding a
randomly chosen particle in a trap of depth $E$ at time $t$ thus
obeys
\be
\label{eqn:trap_eom}
(\partial/\partial t)P(E,t)=-\tau^{-1}(E)P(E,t)+Y(t)\rho(E)
\end{equation}
in which the first (second) term on the RHS represents hops out of
(into) traps of depth $E$, and $Y(t)$ $=$ $\av{\tau^{-1}(E)}_{P(E,t)}$ is
the average hopping rate. The solution of~(\ref{eqn:trap_eom}) for
initial condition $P_0(E)$ is
\be
\label{eqn:trap_solution}
P(E,t)=P_0(E)e^{-{t}/{\tau(E)}} +\rho(E)\int_0^{t}\!\!
dt' \,Y(t') e^{-(t-t')/\tau(E)}
\ee
from which $Y(t)$ has to be determined self-consistently.  For the
specific choice of prior distribution
$\rho(E)\sim\exp\left(-{E}/{\tg}\right)$, the model shows a glass
transition at a temperature $\tg$.  This can be seen as follows.  At a
temperature $T$, the equilibrium Boltzmann state (if it exists) is
$P_{\rm eq}(E)\propto\tau(E)\rho(E)\propto\exp\left({E}/{T}\right)
\exp\left(-{E}/{\tg}\right)$. For temperatures $T\le \tg$ this is
unnormalizable, and cannot exist; the lifetime averaged over the
prior, $\av{\tau}_{\rho}$, is infinite.  Following a quench to
$T\le\tg$, the system never reaches a steady state, but instead ages: in
the limit $\tw\to\infty$, $P(E,\tw)$ is concentrated entirely in traps
of lifetime $\tau=O(\tw)$.  The model thus has just one characteristic
time scale, which grows linearly with the age $\tw$. In contrast, for
$T>\tg$ all relaxation processes occur on time scales $O(\tau_0)$. In what
follows, we rescale all energies such that $\tg=1$, and also set
$\tau_0=1$.

To study FD violation one can extend the model as
follows~\cite{FieSol_condmat}: To each trap we assign, in addition to
its depth $E$, a value for an (arbitrary) observable $m$; by analogy
with spin models we refer to $m$ as magnetization.  The trap
population is then characterized by the joint prior distribution
$\distrm\rho(E)$, where $\distrm$ is the distribution of $m$ across
traps of given fixed energy $E$. We focus on the non-equilibrium
dynamics after a quench at $t=0$ from $T=\infty$ to $T<1$; the initial
condition
is thus
$P_0(E,m)=\distrm\rho(E)$.  The subsequent evolution is governed by
\be
\label{eqn:eomwithmh}
(\partial/\partial t)P(E,m,t)= -\tau^{-1}(E,m)P(E,m,t)+
Y(t)\distrm\rho(E)
\ee
where the activation times are modified, in the presence of a small
field $h$ conjugate to $m$, according to
$\tau(E,m)=\tau(E)\exp\left(mh/T\right)$. Other choices of $\tau(E,m)$
that maintain detailed balance are
possible~\cite{RinMaaBou00,BouDea95}; we adopt this particular one
because, in the spirit of the unperturbed model, it ensures that the
jump rate between any two states depends only on the initial state,
and not the final one.

Let us outline briefly the calculation of correlation and response
function for the observable $m$. The autocorrelation function
is (in the absence of a field, \ie\ $h=0$)
\bea
\CC=\int \!\! dm\ d\mw\ dE\ d\Ew\ m \, \mw \,P(\Ew,\mw,\tw) \times
\nonumber\\
   \times \left[\Emprobc-P(E,m,t)\right]
\label{eqn:correarly}
\end{eqnarray}
in which the `propagator' $\Emprobc$ is the probability that a
particle with magnetization $\mw$ and energy $\Ew$ at time $\tw$
subsequently has $m$ and $E$ at time $t$. The key simplification in
the calculation is that this can be related back to the known
distribution $P(E,m,t)$:
\bea
\Emprobc = \delta(m-\mw)\delta(E-\Ew) e^{-(t-\tw)/\tau(\Ew)}\nonumber\\
        {}+{}\int_{\tw}^{t}\!\!dt' \, \tau^{-1}(\Ew)
e^{-(t'-\tw)/\tau(\Ew)}P(E,t-t')\distrm
\label{eqn:propagator}
\eea
The terms corresponds to the particle not having hopped at all since
$\tw$ (first term) or having first hopped at $t'$ (second term) into
another trap; after a hop the particle evolves as if `reset' to time
zero since it selects its new trap from the prior distribution
$\sigma(m|E)\rho(E)$, which describes the initial state of the system.
We also have $P(E,m,t)=\distrm P(E,t)$ since at zero field the
unperturbed dynamics is recovered. Substituting these relations
into~(\ref{eqn:correarly}) and integrating over $m$ and $\mw$, one
obtains an exact integral expression for $C$. This is expressible as
the sum of two components: The first depends only on the mean
$\trapmean$ of the fixed energy distribution $\distrm$ and the second
only on its variance, $\trapvar$. This of course makes sense, since
the correlation function only measures statistics of $m$ up to second
order.

To find the corresponding response function, one can use the
relation~\cite{FieSol_condmat}
\be
T\deriv{\tw}\RR=\deriv{t}\CC +
\deriv{t}\globalmbr{t}\globalmbr{\tw}
\label{eqn:H3}
\end{equation}
in which $\lav m(t)\rav=\lav\trapmean\rav_{P(E,t)}$ is the global mean
of $m$. Equation~(\ref{eqn:H3}) is a slight generalization of the
results of~\cite{BouDea95,SasNem99}; it is exact for any Markov
process in which the effect of the field on the transition rate
between any two states depends only on the initial state and not the
final one. Substituting the expression for $C$ into~(\ref{eqn:H3}),
and integrating over $\tw$ with the boundary condition
$\chi(\tw,\tw)=0$, one finds a closed expression for $\chi$. For
convenience, we rescale the field $h\to Th$ in the following, thus
absorbing a factor $1/T$ into the definition of the response
function. In this way, the slope of the FD plot of $\chi$ vs $C$, or
of their normalized analogues, becomes $-X=-T/\teff$ ($=-1$ in
equilibrium).

Using the exact results described above, $C$ and $\chi$ have been
calculated for a number of different distributions $\distrm$, each
specified by given functional forms of $\trapmean$ and
$\trapvar$~\cite{FieSol_condmat}. Each form effectively corresponds to
a distinct physical identity of the observable $m$. For example
$\distrm=\delta(m-E)$ implies $m=E$, in which case $h=-\delta
T/T$. For simplicity, we confined ourselves to distributions either of
zero mean (but non-zero variance) or of zero variance (but non-zero
mean). Even for these simpler observables $m$, the equal-time correlator
$C(t,t)$ can depend strongly on $t$; this is in contrast to the
spin models often considered in mean field studies, where $C(t,t)$ is
automatically normalized. The following plots are therefore of the
normalized correlation and response functions, $\Ct$ and $\Rt$; as
explained above, these are obtained by dividing the raw quantities by
the equal-time correlator $C(t,t)$ at the {\em later} time.

For the first class of observables (with zero mean $\trapmean$), we
considered a variance $\trapvar=\exp(En/T)$ for various values of the
exponent $n$, generalizing the results of~\cite{BouDea95} for
$n=0$. The equal-time correlator is then $C(t,t)=\int\! dE\
P(E,t)\exp\left({nE}/{T}\right)$.  For large $t$, its scaling behavior
can be deduced from that of $P(E,t)$, and depends on the value of $n$:
\begin{figure}
\centerline{\psfig{figure=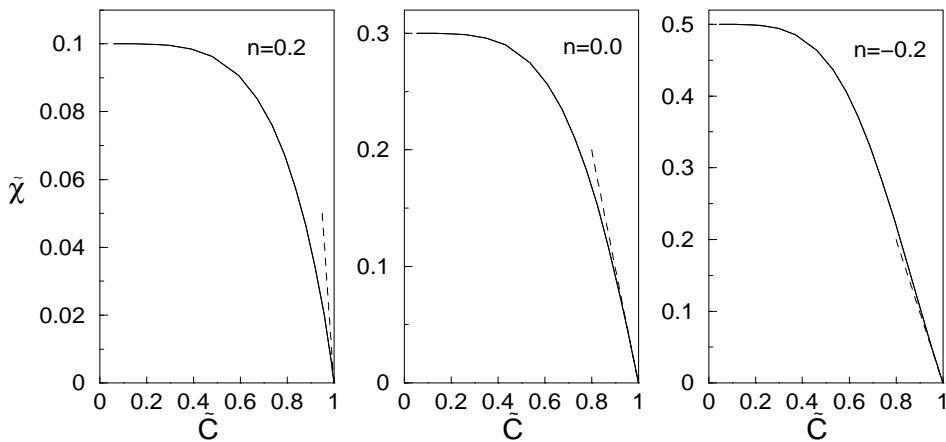,height=6cm,width=12.5cm}}
\caption{
  FD plots of $\tilde{\chi}$ vs $\tilde{C}$ for a distribution
  $\distrm$ of variance $\exp(nE/T)$ (but zero mean) for
  $n=0.2,\,0.0,\,-0.2$; $T=0.3$. For each $n$ data are shown
  for times $t=10^6,\,10^7$; these
  are indistinguishable, confirming that the limiting FD plot has
  been attained. Dashed line: The predicted asymptote
$\tilde{\chi}=1-\tilde{C}$ for
  $\tlim$ and $\tilde{C}\to 1$. From Ref.~\protect\cite{FieSol_condmat}.
\label{fig:var}
}
\end{figure}
For $n<T-1$, $C(t,t)$ is sensitive only to shallow traps (the
population of which depletes in time as $t^{T-1}$ due to ageing,
giving $C(t,t)\sim t^{T-1}$) and one correspondingly expects the
two-time correlator to decay on time scales $\dt=O(1)$, probing only
quasi-equilibrium behavior. In contrast, for $T-1<n<T$ one finds that
$C(t,t)$ is dominated by the contribution from traps with depths $E$
such that $\tau(E)=O(t)$; this results in the scaling $C(t,t)\sim
t^{n}$. For such $n$, the two-time correlator will decay on ageing
time scales $\dt=O(\tw)$, and we expect strong violation of
equilibrium FDT. The numerical results confirm the above
predictions. For values of $n<T-1$ (not shown) the normalized FD plot
approaches a straight line of equilibrium slope $-1$ at long times
$\tlim$. In the regime $T-1<n<T$ (Fig.~\ref{fig:var}), equilibrium FDT
is strongly violated, as expected, but a limiting {\em
non-equilibrium} FD plot is nevertheless approached at long
times. This can be proved analytically by showing that $\tilde C$ and
$\tilde\chi$ share the same scaling variable $(\dt)/\tw$ in this
limit. The slope of each plot varies continuously with $\tilde C$.  In
contrast to mean field, this is not due to an infinite hierarchy of
time sectors; the variation is in fact continuous across the single
time sector $\dt=O(\tw)$. More seriously, different observables give
different plots. One has to be slightly careful here:
Observable-independence requires that $\XX$, as a function of the long
{\em times} $t$ and $\tw$, be independent of the observable considered. The
limiting FD plots, however, give $X$ as a function of $\Ct$ rather
than of $t$ and $\tw$. One thus has to convert the results for $X$
derived from different observables back into the time domain before
comparing them. In our case this means comparing the $X(\Ct)$ values
corresponding to a fixed value $(\dt)/\tw$; such a comparison reveals
that the values of $X$ do indeed differ among the observables
considered.

\begin{figure}
\centerline{\psfig{figure=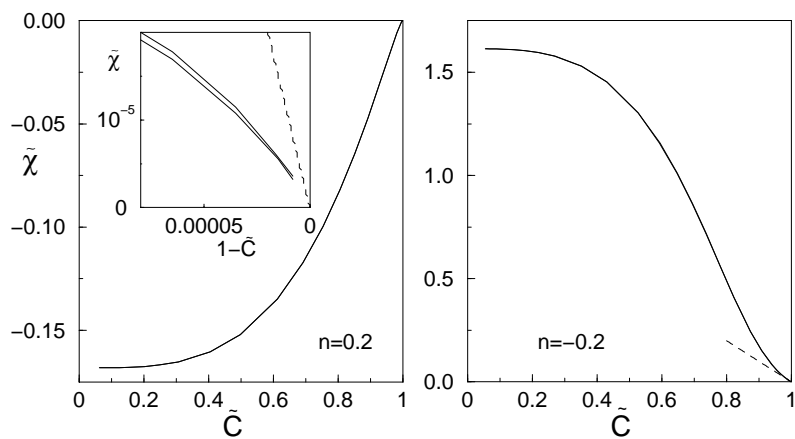,height=6cm}}
\caption{FD plots of $\Rt$ vs $\Ct$ for a distribution with
  mean $\exp(nE/2T)$ (but zero variance), for $n=0.2,\,-0.2$; $T=0.3$.
  Curves are shown for times $t=10^6,\,10^7$,
  but are indistinguishable except for the zoom-inset on the left hand
(upper curve: $t=10^7$). Dashed line: The predicted asymptote
$\tilde{\chi}=1-\tilde{C}$ for $\tlim$, $\tilde{C}\to 1$. 
From Ref.~\protect\cite{FieSol_condmat}.
\label{fig:mean}
}
\end{figure}

In Ref.~\cite{FieSol_condmat}, we also considered observables defined
by a fixed energy distribution $\distrm$ with zero variance and mean
$\trapmean=\exp(En/2T)$; in this case $m$ is a deterministic functions
of $E$. Here too, we found that for $n<T-1$ the limiting FD plot is of
equilibrium form, while for $T-1<n<T$ (Fig.~\ref{fig:mean}) a
non-equilibrium FD plot is approached as $\tlim$, with a shape
dependent (here very obviously) on the observable $m$. Note that the
response function can now change sign from positive to negative for
increasing time separation $t-\tw$. This is not untypical for
acticated dynamics, and can be understood as follows: Suppose for
simplicity that $n>0$. Then $\trapmean$ increases with $E$ and so in
equilibrium the application of a positive field conjugate to $m$ would
increase $E$ and thus $m$ on average, giving the conventional {\em positive}
response which we observe for $t$ close to $\tw$. On the other hand,
since the application of the field effectively increases the depth of
all traps it also slows down the out-of-equilibrium relaxation of $E$
and $m$ to larger values, thus giving a {\em negative} response for large
$t-\tw$.

Observables with a power law dependence of the mean, $\trapmean=E^p$,
were also considered, with zero variance $\trapvar=0$ as before so
that $m=\trapmean$ is a deterministic function of $E$.  Then $p=1$
corresponds to energy-temperature FD, since the observable is energy
($m=E$) and the conjugate field is, apart from trivial scale factors,
the temperature.  Interestingly, in this case no limiting plot
exists. This is
because the amplitude of the correlator remains finite as $\tlim$,
while the response function, for any fixed value of $\tilde{C}$,
diverges as $\ln t$.

A final observation to be made from Figs.~\ref{fig:var}
and~\ref{fig:mean} is that, whenever a limiting FD plot exists for the
observables considered here, its slope in the limit of well separated
times ($t\gg\tw$, corresponding to $\Ct(t,\tw)\to 0$) is always
$X_\infty = 0$.

\section{Model system 2: Glauber-Ising chain at zero temperature}

As the second model system, we consider a chain of $N$ spins $s_i$
with the Ising Hamiltonian $H=-J\sum_i s_i s_{i+1}$. The time
evolution is assumed to be given by Glauber dynamics; this means that
at each time step a spin $i$ is chosen randomly and its state assigned
to $s_i=\pm 1$ with probability proportional to the Boltzmann factor
$\exp(- H/T)$ for the new configuration. Time is scaled so that each
spin is updated once, on average, per unit time.

When this system is quenched to a low temperature from a random
initial state, corresponding to equilibrium at $T=\infty$,
the dynamics will progressively line up neighbouring spins with each
other and alternating domains of up and down spins will grow (coarsen)
until they reach the equilibrium domain length $\xi_{\rm eq} \sim
\exp(2J/T)$ at a time $\tau_{\rm eq} \sim \exp(4J/T)$. In the limit of
$T\to 0$, these scales grow to infinity and the systems remains out of
equilibrium at any finite time; this will be the limit of interest to
us. Since the Ising spin chain has a phase transition to an ordered
ferromagnetic state at $T=0$, one can regard the out-of-equilibrium
dynamics for $T\to 0$ as critical
coarsening~\cite{GodLuc_ferro}. Physically, however, this is somewhat
different from coarsening at a finite-temperature critical point,
since in the Ising chain the growing domains are not critical but have
the structure of either of the ordered phases (spins all up or all down).

The FD relations for the Glauber-Ising chain have been studied
previously for the case of the spin-spin autocorrelation (see
Ref.~\cite{GodLuc00}; a brief summary of the results can be found in
Ref.~\cite{GodLuc_ferro}).  It is a
simple matter to extend the calculation to a somewhat more general
class of observables, defined as
\be
m = \sum_i \epsilon_i s_i
\label{m_def}
\ee
where the $\epsilon_i$ are quenched random variables with zero mean
and translation-invariant covariances
$[\epsilon_i\epsilon_j]=q_{j-i}$; here $[\ldots]$ denotes the average
over the quenched disorder. The corresponding perturbation in the
Hamiltonian is $-hm$, as usual, so that the local field acting on each
site is $h\epsilon_i$. With different choices of the correlation
function one can then interpolate between the two extremes of a
uniform field ($q_k=1$, in which case $m$ is the conventional
magnetization) and a locally random field ($q_k=\delta_{k0}$, the case
treated in Ref.~\cite{GodLuc00}).

Because $\lav s_i(t)\rav=0$, $m$ as defined by\eq{m_def} also has an
average of zero at all times. Using the known results for the two-time
spin-spin correlation function~\cite{GodLuc00}, $C_{ij}(t,\tw) = \lav
s_i(t)s_j(\tw)\rav$, which can be written as $C_{ij}(t,\tw) =
c_{j-i}(t,\tw)$ due to translation invariance, one then has for the
autocorrelation function of $m$ (scaled by $N$ to get a quantity of
order unity)
\be
\fl C(t,\tw) = \frac{1}{N}\sum_{ij} \epsilon_i \epsilon_j c_{j-i}(t,\tw) =
\sum_k \left(\frac{1}{N}\sum_l \epsilon_{l}\epsilon_{l+k}\right)
c_k(t,\tw) = \sum_k q_k c_k(t,\tw)
\label{C_GI}
\ee
where in the last step we have used that in the limit of an infinitely
large system the spatial average of $\epsilon_{l}\epsilon_{l+k}$
becomes identical to the quenched average
$[\epsilon_{l}\epsilon_{l+k}]=q_k$. The scaled response function of
$m$ to its conjugate field is obtained similarly as $R(t,\tw) = \sum_k
q_k r_k(t,\tw)$, where the translationally invariant response function
$r_{j-i}(t,\tw)$ gives the response of a spin at site $i$ and time $t$
to a local field applied at site $j$ at time $\tw$. Using the exact
solutions for $c_k(t,\tw)$ and $r_k(t,\tw)$, one can thus evaluate
the correlation and response functions of $m$, at least in principle.

Before giving the results, let us briefly discuss what behaviour one
would expect. We can define a correlation length of the random fields
$h_i=h\epsilon_i$ as $\xih=\sum_{k=-\infty}^\infty q_k$. This length
needs to be compared with the growing spin-spin correlation length
$\xi(t)$ of the coarsening system, which is of the same order as the
typical domain length. For early times the field is essentially
uniform on the lengthscale of the spin correlations ($\xi(t)\ll\xih$)
and so we expect to see behaviour corresponding to the uniform field
limit. For late times, the field correlations are much more
short-ranged than those of the spins ($\xi(t)\gg\xih$); across each
typical spin domain, the $\epsilon_i$ change signs many times and we
expect qualitatively the same effects as for completely uncorrelated
fields. The crossover between these two regimes takes place when
$\xi(t)\approx \xih$.  Since the growth of $\xi(t)$ is driven by the
random-walk motion of the domain walls, it scales as $\xi(t)\sim
t^{1/2}$, and thus the crossover time is of order $\tauh=\xih^2$.

To simplify the theoretical analysis, it is convenient to look at the
limit where all times ($t$, $\tw$ and $\tauh$) are large compared to
unity. This then also implies that the field correlation length $\xih$
is large, so that we can write $q_k=q(k/\xih)$, where $q(x)$ is a
smooth function of the scaled distance $x=k/\xih$. We leave all
details of the calculation to a forthcoming publication and display
only the results.

\begin{figure}
\epsfig{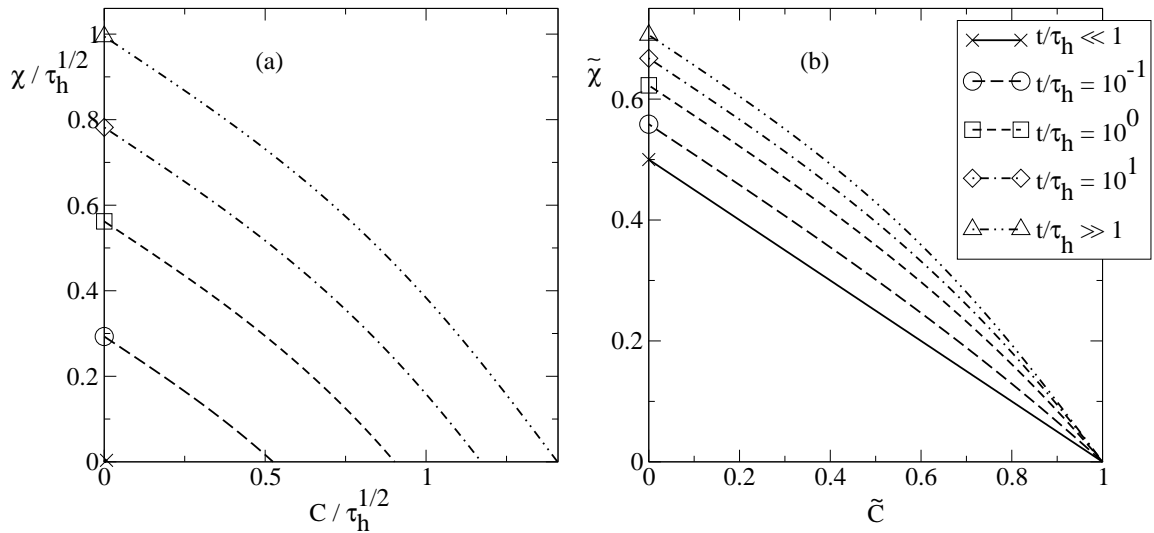}
\caption{FD plots for Glauber-Ising chain, for a linear
observable~(\protect\ref{m_def}) with fields $\epsilon_i$ that have a
Lorentzian covariance function $q_k=q(k/\xih)$ with $q(x)=1/(1+x^2)$.
(a) Raw, unnormalized, response and correlation functions. The
constant factor $1/\tauh^{1/2}$ only serves to remove the trivial
scaling of the amplitude of these functions with the field correlation
length $\xih=\tauh^{1/2}$. Curves are for increasing ratios of
$t/\tauh$ as shown in the legend. (b) FD plots of the normalized
correlation and response, $\Ct$ and $\Rt$. The normalization shows
that in the limit $t/\tauh\ll 1$ one obtains a straight line of
negative slope $X=1/2$, exactly as in the long time limit for {\em
uniform} fields; see Fig.~\protect\ref{fig:limits}.
%
%
\label{fig:Lorenz}
}
\end{figure}
In Fig.~\ref{fig:Lorenz}(a) we show the unnormalized (except for a
constant scale factor) FD plot for the case of a Lorentzian field
correlation function $q(x)=1/(1+x^2)$. As before, the earlier time
$\tw$ is used as the curve parameter and $t$ is held constant for each
curve. As $t$ increases, the amplitude of both correlator and response
initially grows, but eventually saturates for $t\gg\tauh$. For such
long times, the FD plot is identical to that for uncorrelated
fields~\cite{GodLuc00} as predicted, with a negative slope $X=1$ for
$t$ close to $\tw$, corresponding to quasi-equilibrium, and
$X=X_\infty=1/2$ in the limit of well separated times $t\gg\tw$. For shorter
times $t\ll\tauh$, on the other hand, the FD plot is a straight line
of slope $X=1/2$ throughout; this is more easily seen once correlation
and response are normalized by $C(t,t)$ as in
Fig.~\ref{fig:Lorenz}(b). As we will see very shortly, and consistent
with our expectations, the FD plot in this regime, \ie\ before the
crossover at $t\approx\tauh$, is identical to that obtained for
uniform fields.

\begin{figure}
\begin{center}
\epsfig{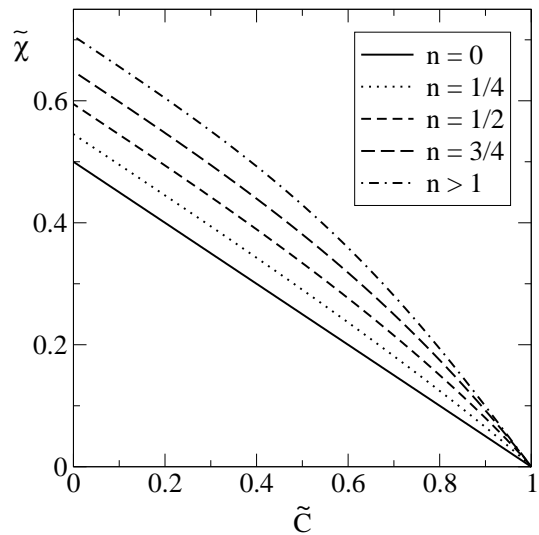}
\end{center}
\caption{Limiting FD plots for the Glauber-Ising chain, for different
asymptotic behaviours $q(x)\sim x^{-n}$ of the field correlation
function; the value of $n$ is shown in the legend. For $n\to 0$ one
has the case of uniform fields, which gives a straight line of
negative slope $X=1/2$. The opposite extreme $n\geq 1$ corresponds to
correlation functions with a finite correlation length; these all give
the same limit plot, identical to that for entirely uncorrelated
fields. Values of $n$ between 0 and 1 give nontrivial intermediate FD
plots. Note that the negative slope $X$ of all plots is, in the limit
of well separated times, $X_\infty=1/2$.
\label{fig:limits}
}
\end{figure}
Fig.~\ref{fig:limits} shows the limiting ($t\to\infty$) FD
plots for a range of field correlation functions $q(x)$. At one
extreme, we have the case of uniform fields ($q(x)=1$), giving a
straight line of negative slope $X=1/2$. At the other extreme, all
field correlation functions $q(x)$ for which a finite correlation
length can be defined have the same limit plot as for uncorrelated
fields. Because we defined the correlation length above as
$\xih=\sum_{k=-\infty}^\infty q_k$, the scaled correlation function must
obey $\int\!dx\,q(x)<\infty$ for a finite correlation length to exist;
this requires that $q(x)$ decay faster than $x^{-1}$ for large
$x$. Correlation functions $q(x)$ which decay asymptotically as
$q(x)\sim x^{-n}$ with $0<n<1$ give intermediate FD plots, whose shape
is determined only by the asymptotic decay exponent $n$. Results for
several values of $n$ between 0 and 1 are shown in
Fig.~\ref{fig:limits}; they demonstrate clearly that even among the
relatively restricted class of observables defined by\eq{m_def},
nontrivially different limiting FD plots exist which give different
results for the FD ratio $X$. Intriguingly, however, the limit of $X$
for strongly separated times $t\gg\tw$, is $X_\infty=1/2$ for {\em
all} observables considered here.

\section{Discussion}
\label{sec:discussion}

Let us summarize our findings. For the trap model, on the basis of
exact relations for correlation and response functions, we have
studied FD relations in the glass phase for a broad class of
observables $m$. For observables which indeed probe the ageing regime,
equilibrium FDT is violated; in most cases, a limiting non-equilibrium
FD plot is approached at long times. In contrast to mean field, this
plot depends strongly on the observable. It furthermore has a slope
which varies continuously across the single time sector
$\dt=O(\tw)$. This shows that in this simple, paradigmatic model of
glassy dynamics, the mean field concept of an FD-derived effective
temperature $\teff$ cannot be applied. One may dismiss the trap model
as too abstract for this to be of general relevance; but by placing
traps onto a $d$-dimensional lattice and allowing hops only to
neighbouring traps, a more `physical' model (of diffusion in the
presence of disorder) can be obtained. Recent work~\cite{RinMaaBou00}
shows the scaling of correlation functions to be unaffected by this
modification; since~(\ref{eqn:H3}) would also continue to hold, the
results for the trap model should be qualitatively unchanged.

Our results for the Glauber-Ising chain were qualitatively
similar. There, we restricted ourselves to observables that are linear
in the spin variables; again, exact results for the correlation and
response functions of such observables can be obtained. Limiting FD
plots always exist, but depend nontrivially on the observable.  Except
for the case where the observable is the magnetization (and the
corresponding perturbation therefore a uniform field), the slope of
the limit plots varies continuously; as in the trap model, this
variation takes place within a single time sector $\dt=O(\tw)$.

The results above demonstrate that the idea of an effective
temperature $\teff$ derived from FD relations does not apply
straightforwardly to generic non-mean field models. Could the concept
nevertheless be rescued? One difficulty in the two models considered
was the non-uniqueness of the limiting FD plots. One could argue that
in order to probe an inherent $\teff$ characterizing the
non-equilibrium dynamics, the statistical properties of the observable
must not change significantly across the phase space regions visited
during ageing.  (In coarsening models, similar arguments have been
used to exclude observables correlated with the order
parameter~\cite{Barrat98}.)  Applying this idea to the trap model,
where the typical trap depth $E$ increases without bound for
$t\to\infty$, a `neutral' observable would presumably require that
$\trapmean$, $\trapvar\to$ const.\ as $E\to\infty$; with this
restriction we indeed get a unique limiting FD plot. In the
Glauber-Ising chain, the energy $E=\lav H\rav$ decreases towards the
ground state energy $E_0$ during coarsening. A neutral observable
might then be one whose properties within a {\em microcanonical}
ensemble at given $E$ do not change significantly as $E\to E_0$. It is
easy to see that, for our observables\eq{m_def}, this criterion again
produces a limiting FD plot: It only allows field correlation
functions with a finite correlation length---which all give the same
limit plot, namely that for uncorrelated fields---while other
observables have a variance which diverges as $E\to E_0$. The
selection of neutral observables may in fact obviate the normalization
problem for FD plots discussed in detail in this paper: If an
observable has a variance $C(t,t)$ which either converges to zero or
diverges as $t$ increases, then this in itself implies that its
statistical properties change strongly while the system ages;
conversely, for neutral observables $C(t,t)$ should approach a finite
limit for $t\to\infty$.

Even if one can, by a judicuous choice of a `neutral' observable,
produce a unique limiting FD plot and thus a unique FD ratio
$X(t,\tw)$, one is still faced with the problem that $X$ may---as in
our cases---vary continuously across a single time-sector. This almost
certainly makes $X$ inappropriate for the definition of a meaningful
effective temperature $\teff=T/X$, since two thermometers probing time
scales which differ only by factors of order unity should be required
to measure the same temperature. There is, however, the intriguing
possibility that the limit of $X$ obtained for long times and large
time separations, $X_\infty=\lim_{\tw\to\infty}\lim_{t\to\infty} \XX$
may still correspond to a meaningful $\teff$. In fact, in the two
models considered here, $X_\infty$ was the same for all observables
considered, as long as limiting FD plots actually existed. These ideas
should be explored in future work; we are currently in the process of
calculating FD plots for {\em second order} observables such as
$m=\sum_i\epsilon_i s_i s_{i+1}$ in the Glauber-Ising chain. If the
same $X_\infty$ results, this would strongly support the hypothesis
that $X_\infty$ is indeed observable-independent.

Finally, it is of course likely that there are different classes of
non-mean field systems as far as the concept of an effective
temperature is concerned. One could conjecture that FD plots would
provide a diagnostic tool here: The appearance of rounded FD plots
like those shown above may signal that a system does not have a
meaningful $\teff$, while limiting FD plots consisting of two straight
line segments---as found in simulations of some non-mean field
models~\cite{KobBar00,AreRicSta00,Barrat98}---may indicate that a
well-defined effective temperature does indeed exist. One of the main
tasks ahead must be to develop physical criteria that allow us to
predict to which of these two classes a given system belongs. It is
probable that a strong separation of the dynamics into fast and slow
time scales will play a role (see \eg~\cite{LeuNie01_proc,Garriga01}),
but this may
not be sufficient.  The trap model is again instructive here: In the
glass phase the fast processes (on time scales of order unity) do not
contribute to the dynamics, and all relevant relaxation processes
happen on slow time scales of the same order ($\sim\tw$). A naive
time scale separation argument would then suggest that the FD plot
should consist of a single straight line, with a non-equilibrium slope
$X\neq 1$, while the exact results presented above show that the
actual situation is rather more complicated.

{\bf Acknowledgements}: We thank L.\ Berthier, J.\ P.\ Bouchaud, M.\
E.\ Cates, M.\ R.\ Evans, J.-P.\ Garrahan and F.\ Ritort for
helpful suggestions. We also acknowledge financial support from EPSRC
(SMF) and the Nuffield Foundation (PS, grant NAL/00361/A).

\section*{References}

\bibliography{/home/psollich/foams/references,local}

\end{document}